\documentclass[12pt,fleqn]{article}

\usepackage{latexsym}
\usepackage{epsfig}
\usepackage{amsmath}

\newcommand{\enproof} {\hfill $\Box$ \bigskip}

\newtheorem{theorem}{{\bf Theorem}}[section]
\newtheorem{lemma}{{\bf Lemma}}[section]

\begin{document}

\title{Correction to: A Practical, Provably Linear Time, In-place and Stable Merge Algorithm via 
the Perfect Shuffle
\thanks{This work was supported in part by the Natural Sciences and Engineering
        Research Council of Canada}}
\author{{\sc John Ellis$^\dagger$} and { \sc Ulrike Stege$^\dagger$}  \\
\ \ \\
     {\small \em $^\dagger$Department of Computer Science} \\
     {\small \em University of Victoria, Canada}}
%     {\small \em Victoria, British Columbia, V8W 3P6, Canada}}  
\maketitle
\begin{abstract}
	We correct a paper previously submitted to CoRR.
	That paper claimed that the algorithm there described was provably of linear time
	complexity in the average case.  The alleged proof of that statement contained an
	error, being based on an invalid assumption, and is invalid.  In this paper we present 
	both experimental and analytical
	evidence that the time complexity is of order $N^2$ in the average case, where $N$ 
	is the total length of the merged sequences.
\end{abstract}
\noindent
\textbf{keywords:} algorithm, perfect shuffle, merging, sorting, stability, in-place

\section{Introduction}
We correct an invalid claim made in \cite{EllSteg:2015}.
That paper described an in-place, stable merge algorithm
and presented a ``proof'' of average case, linear time complexity.
That alleged proof was invalid, being based on an incorrect assumption.
In this paper we point to the error in the proof and offer an alternative
analysis of the time complexity.
We also report experimental results of executing the algorithm.
These results are consistent with time complexity of order $n^2$, in the average case.

\section{Experimental Results} 
\label{exper}
\begin{figure}
\begin{center}
\epsfig{file=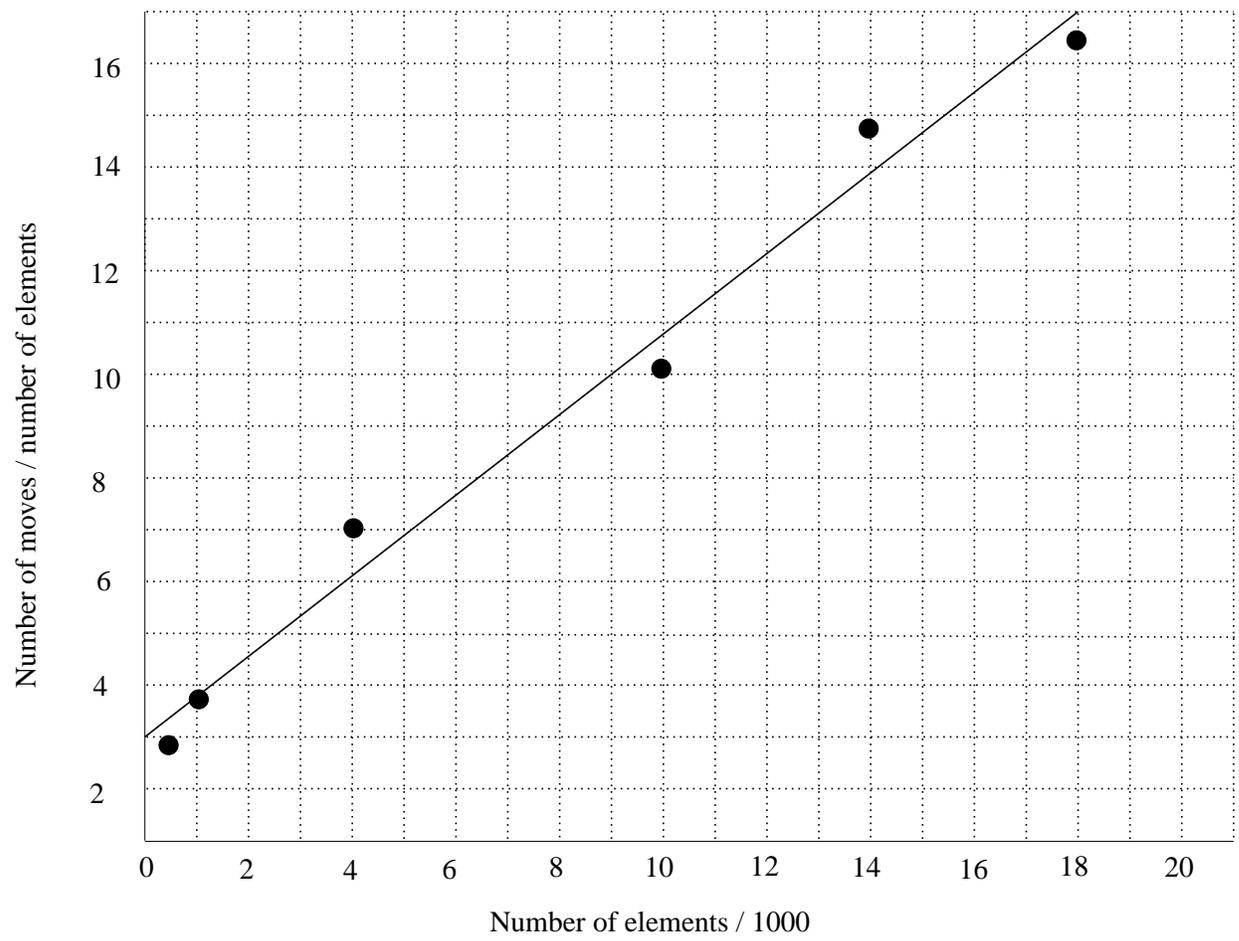, width=6.5in}
\end{center}
\caption{Number of moves per element}
\label{exp}
\end{figure}

No experimental results were reported in \cite{EllSteg:2015}, except for citing
\cite{Dalkilic:2013} where some results of experiments  were reported  that were
consistent with linear time.
Those experiments were conducted on quite long sequences of integers.
In \cite{Dalkilic:2013} Table 1 the sizes go from $2^9$ to $2^{29}$
and in Figure 3 they go up to about 50 million.
What may be very relevant is the statement that the numbers being processed were 8-bit 
integers, i.e., much smaller that the sequence length, so that the sequences being 
merged comprise a relatively small number of long sub-sequences of identical numbers.
The performance of the algorithm on these very special distributions says little
about its expected performance on more general distributions.
This may explain the discrepancy between these results and those of our own which
we now present.

The elements of our integer data sets were drawn at random from 1 through $4N$, where $N$ is the
combined lengths of the merged sequences, which to some extent minimised the occurrence of
duplicate elements.
We experimented only with input lists of equal length $N/2$.
For each length we ran 10 tests counting the number of element moves and comparisons, as
a measure of the time taken.
The average number of moves over the 10 tests divided by $N$ is plotted vs. $N$ over the range 500 - 20000
in Figure \ref{exp}.
The ratio $moves/N$ grows approximately linearly with $N$ which indicates that the
number of moves itself is of order $N^2$.
Interestingly, the number of comparisons per element is consistently around 1.0, so a plot was
not necessary.
A worst case upper bound of $O(N^2)$ was established in \cite{EllSteg:2015}.
Such slow behaviour makes the algorithm, as it stands, of little practical value. 

\begin{figure}
\begin{center}
\epsfig{file=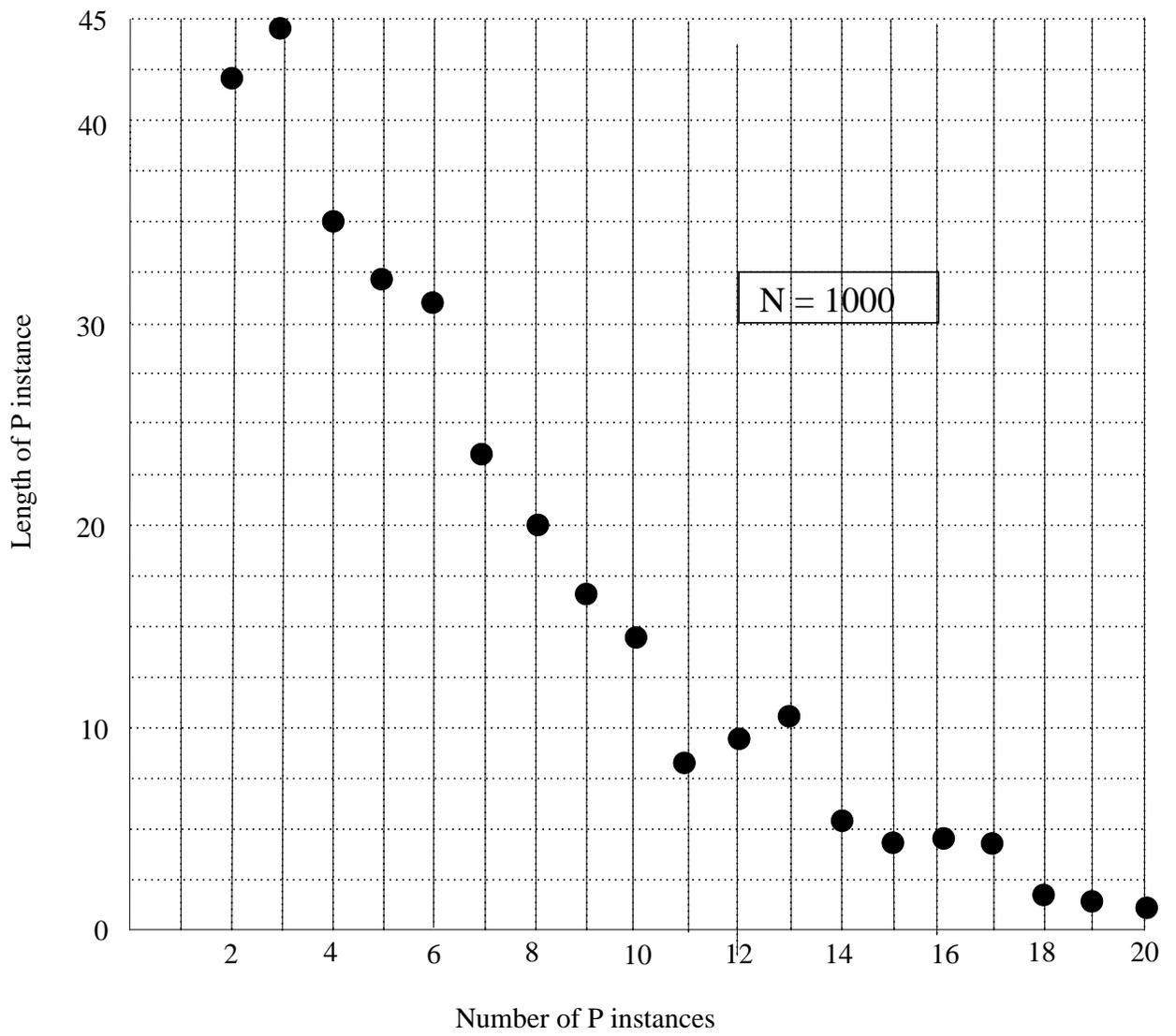, width=6.5in}
\end{center}
\caption{Frequency of instances of $P$ of a certain length}
\label{freq1}
\end{figure}

\begin{figure}
\begin{center}
\epsfig{file=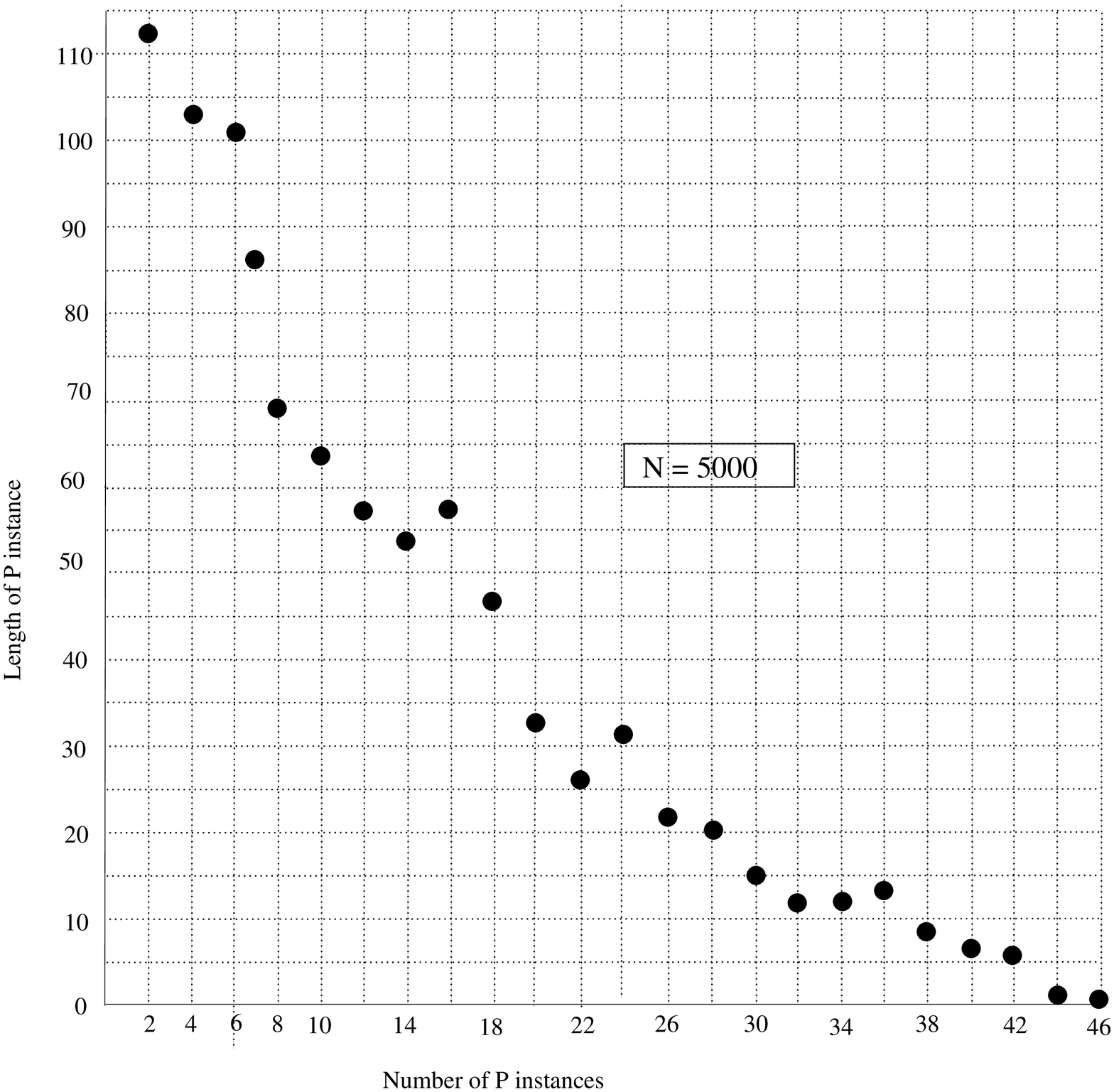, width=6.5in}
\end{center}
\caption{Frequency of instances of $P$ of a certain length} 
\label{freq2}
\end{figure}

We also recorded the length of the $P$ segments (see Appendix 1) at the time they were rotated and the frequency of 
occurrence of segments of each size.
The results are displayed in Figures \ref{freq1} and \ref{freq2}.
These plots show an increase in the frequencies of $P$ segments of a particular length as $N$ increases, which 
is consistent with the super-linear time performance.

\section {Errors in the previous presentation}
\label{err}
We reproduce the algorithm from \cite{EllSteg:2015} in Appendix 1.
The alleged proof of linear time is reproduced in Appendix 2.
In each traversal of the outer loop in the algorithm 
elements are moved either by a single exchange or by a shuffle or by a rotation.
Each rotation  requires moves $\geq c(|P| + |D| + |O|)$ for some constant $c$.
See Figure \ref{sublist1}.
In Appendix 2, Lemmas 1 and 2 attempt
to show that the probability of the occurrence of a $P$ sequence of a particular length 
decreases exponentially with the length of the sequence.
From this Lemmas 3, 4 and 5 deduce that the expected time spent scanning, shuffling or rotating
is constant and independent of the input size.
Finally Theorem 1 uses these Lemmas to show that the average time complexity is $O(N)$
where $N$ is the total length of the input.

This result is not supported by the experimental results just presented in Section \ref{exper}.
The reason is that the proofs of Lemmas 1 and 2 are based on a false assumption.
Consider Lemma 2 regarding $|P|$ which grows with increasing $N$ according to the experiments.
The proof is based on the assumption that the probability that $|P| = p$
is given by the number of possible arrangements of remaining unplaced elements
where $|P| = p$ divided by the number of all possible arrangements of the remaining
unplaced elements.
This is valid only if the elements in $P$ and $Sh$ are randomly, independently and identically
distributed.
This may be true at the start of the process, as is the case for our experiments where the 
input lists are pseudo-randomly chosen,
but does not necessarily remain true as the algorithm progresses
because elements in $O$ have been selected to be less than the first element in $P$ before insertion 
in front of $P$.
Hence the proof is not valid.

\section{Time Complexity Analysis}

%We construct a function of $N$ that describes the probability that $|P|$ reaches a certain size.
%We note that, Figure \ref{sublist1}, at the point of rotation of the $P-O$ sequence the first element 
%in $O$ is less than the first element in $P$.
%The rotation requires at least $|P|$ moves.
We construct a function of $N$ that describes the probability that the $n^{th}$ element in one
of the merged sequences is less than the $(n-k)^{th}$ element in the other.
Then we examine this function experimentally. 

Let $X_1,X_2,\ldots,X_N$ denote a set of integers selected independently and identically distributed from
$(1,2,\ldots,4N)$, where the "4" can be any integer $>N$.
Let the ordered sequence be denoted by $X_{(1)},X_{(2)},\ldots,X_{(N)}$, so that 
$X_{(1)} \leq X_{(2)} \ldots \leq X_{(N)}$. \\
Let $Y_1,Y_2,\ldots,Y_N$ denote a set of integers selected independently and identically distributed from
$(1,2,\ldots,4N)$, where the "4" can be any integer $>N$.
Let the ordered sequence be denoted by $Y_{(1)},Y_{(2)},\ldots,Y_{(N)}$, so that
$Y_{(1)} \leq Y_{(2)} \ldots \leq Y_{(N)}$.\\
Let $M = 4N$ and let $P(statement)$ denote the probability that the \emph{statement} is true.

Then, from basic order statistics, we have the following probability distributions:
\begin{description}
\item[(a)]
$P(X_{(1)} \leq x) = 1-(1-x/M)^N$, $x = 1,2,\ldots,M $.
\item[(b)]
$P(X_1 \leq x) = x/M$, $x = 1,2,\ldots,M $, which function is denoted by $F(x)$, the cumulative density
function.
\item[(c)]
\[P(X_{(n)}) \leq x) = \sum^N_{j=k} \left(\begin {array}{c} N\\j \end{array} \right) F(x)^j (1-F(x))^{N-j} \]
 $x = 1,2,\ldots,M $, $n = 1,2,\ldots,N$, which function is denoted $G_n(x)$. 
\item[(d)]
For a given $k$, $k = (0,1,2,\ldots)$, $n-k \geq 1$, $n \leq N$,\\
\[
P(X_{n-k} > Y_n)  =  \sum^M_{x=1}P(X_{(n-k)} > x)\cdot P(Y_{(n)} = x)  
                  = \sum^M_{x=1} (1-G_{n-k}(x)) \cdot (G_n(x) - G_n(x-1))
\]
where  $G_n(0) = 0$. 
\end{description}

\begin{figure}
\begin{center}
\epsfig{file=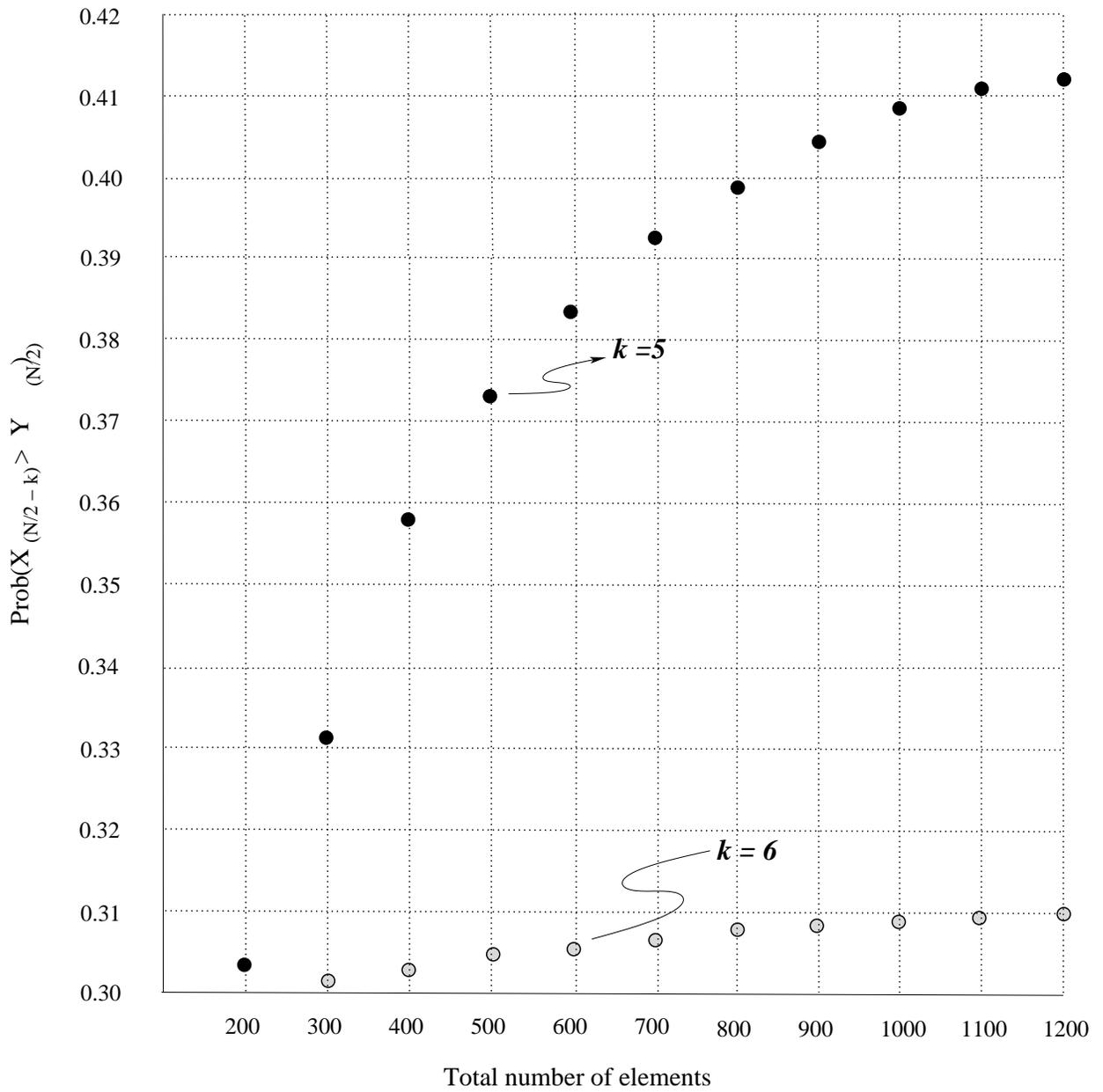, width=6.5in}
\end{center}
\caption{$P(X_{n-k} > Y_n)$} 
\label{prob}
\end{figure}

We computed the value of $P(X_{n-k} > Y_n)$ for various values of $n,N$ and $k$.
The results are displayed in Figure \ref{prob}.
We note that:
\begin{enumerate}
\item
$P(X_{n-k} > Y_n)> k$, for a particular $k$, increases with $N$.
\item
$P(X_{n-k} > Y_n) > k$, for particular $N$, decreases with increasing $k$.
\end{enumerate}

These observations are consistent with the experimental results displayed in Figures \ref{freq1} and
\ref{freq2}.

If the $n^{th}$ element in one of the merged sequences is less than the $(n-k)^{th}$ 
element in the other, then there exists a $D$ segment in the shuffled sequences of length $\geq 2k$.
See Figure \ref{sublist1}.
At some point the algorithm is required to unpack this segment and rotate the $P-O$ segment, where
$|P| \geq 1$.
This  requires at least $|D| + |O| + 1$ moves, since $|P| \geq 1$, but can not guarantee no more than 
$|O| + 1$ elements are now in their final locations.
That is, $\geq 3k+1$ moves are needed to place $k+1$ elements.
Since the frequency of occurrence of $D$ segments of length $2k$ increases with $N$ this causes the 
number of moves to be super-linear in $N$.

\section{Conclusions}

We have presented experimental results and a theoretical analysis showing that
the time complexity of this algorithm is super-linear in $N$, even in the average case.
Consequently, it does not improve on the original perfect shuffle based algorithm \cite{EllMar00}.
We suggest that it might be interesting to consider if the initial shuffling of the merged sequences
necessarily introduces so much disorder that, on the average, the 2-ordered sequence can not be
sorted in linear time. 

\vspace {4mm}
\noindent
\textbf{Acknowledgement}
The authors wish to thank Julie Zhou for showing us how to use order statistics.

\bibliography{sort-merge}
\bibliographystyle{plain}

\newpage
\Large \textbf{Appendix 1:}  The algorithm for equal length lists
\normalsize
\vspace{4mm}
\label{algo}
\begin{figure}
\begin{center}
\epsfig{file=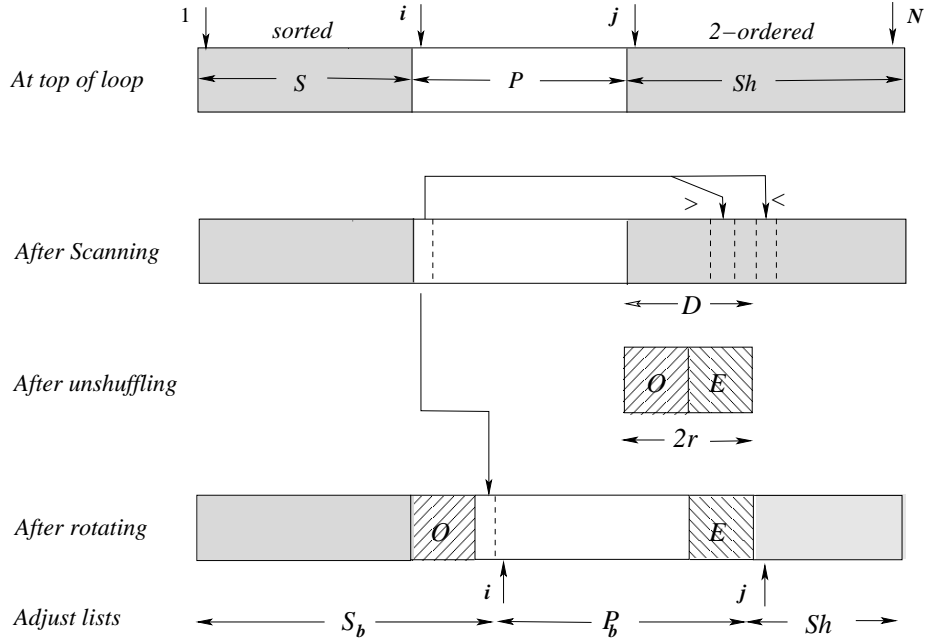, width=4.8in}
\end{center}
\caption{The three sub-arrays}
\label{sublist1}
\end{figure}

Suppose the lists are of equal length.
The process maintains three lists:
a sorted list, $S$, an intermediate  list, $P$ and a 2-ordered list, $Sh$.
Figure \ref{sublist1} illustrates these structures.
The array indices $i$ and $j$ are used to delineate the extents of the three lists.
Index $i$ defines the beginning of $P$ and $j$ the beginning of $Sh$. 
The list $S$ comprises $A[1]$ - $A[i-1]$, $P$ is $A[i]$ - $A[j-1]$ and $Sh$ is $A[j]$ - $A[N]$.

The algorithm Right-going-merge, see Algorithm 1, uses four procedures.
The procedure Scan returns an integer $r$ such that $A[j]$ - $A[j+2r-1]$ is a maximal, even length prefix of 
$Sh$, denoted $D$, such that all odd indexed elements are less than $A[i]$, the first element of $P$.
Scan is only invoked if $|Sh| \geq 2$.
The procedure Shuffle performs an in-shuffle on the input lists, assumed to be of equal length,
i.e., only the interior elements are moved, the first and last elements are left unmoved.
The procedure Unshuffle performs the inverse of Shuffle, i.e., an un-in-shuffle, on $D$ to produce the two lists 
$O$ and $E$. 
Shuffling methods are discussed in Section 4.
The procedure Rotate circularly shifts the two adjacent segments of $A$ that represent $P$ and $O$
to the right by $r$.
See Figure \ref{sublist1}.
We call this procedure \emph{right-going-merge} because the scan proceeds from left to right.
As described in Section 2.3, to handle the case where the lists are not of equal length, we
also use the mirror image of this procedure, called the \emph{left-going-merge}, which scans
from right to left.
\newpage
\begin{tabbing}
1234\=1234\=1234\=1234\=12345\=12345\=12345\=12345         \kill

\>Create $Sh$ by applying Shuffle to the two, equal length lists;\\
\>\{Recall that $A[i]$ is $P[1]$ and $A[j]$ is $Sh[1]$\}\\
\>$i :=$ index of first element in $Sh$;  $j := i+1$;\\
  \\
\>\textbf{while} \textbf{not} $Sh$ is empty \textbf{do}\\ 
\>\>\textbf{if} $P[1] < Sh[1]$ \textbf{then} \{adjust lists\} i := i+1\\
\>\>\>\textbf{if} $|P| = 0$ \textbf{then} $j := j + 1$; complement(\emph{type}) \textbf{fi}\\
\>\>\textbf{else if} $|Sh| = 1$ \textbf{then} $r$ := 1; Rotate; \{adjust lists\} i := i+1; j:=j+1 \\
\>\>\textbf{else} \{Figure 1\} Scan; Unshuffle; Rotate; \\
\>\>\>\{Adjust lists\} $i := i+r+1;$  $j := j+2r$ \textbf{fi} \textbf{fi}\\
\>\textbf{endwhile};
\end{tabbing}
\begin{center}
Algorithm 1: Right-going-merge
\end{center}

\newpage
\noindent
\Large \textbf{Appendix 2:} The invalid proof
\normalsize

\begin{lemma}
\label{Prscan}
If $r>0$ then $Pr(|D| = 2r) \leq 1/2^r$. 
\end{lemma}
\noindent
\textbf{Proof}
\ \ \\
Let the number of elements in $P$, which are all from one list, say $L$, plus the number of $L$
elements in $Sh$, be $n$ and the number of $R$ elements in $Sh$ be $m$.
The number of possible merged arrangements of these $n+m$ elements is $(n+m)!/(n! m!)$.

Suppose Scan defines $D$ such that $|D| = 2r > 0$.
We note that $m\geq r$ and $n \geq m$.
After the Unshuffle, Rotate and redefinition of the list, the number of $L$ elements in $P$ and 
$Sh$ is $n-1$ and the number of $R$ elements in $Sh$ is $m-r$.
See Figure \ref{sublist1}.
The number of arrangements consistent with this fact is $(n+m-r-1)! / ((n-1)!(m-r)!)$.
Hence the probability that  $|D| = 2r$ is given by:

\begin{eqnarray}
Pr(|D| = 2r) &=& \frac{(n+m-r-1)! n! m!} {(m-r)! (n-1)! (n+m)!} \nonumber \\
      &=& \frac{n.m(m-1) \cdots (m-r+1)}{(n+m)(n+m-1) \cdots (n+m-r)} \nonumber \\ 
      &=& \frac{n}{n+m-r} \times \frac{m}{n+m} \times \frac{m-1}{n+m-1} \cdots \times \frac{m-r+1}{n+m-r+1} \nonumber \\
      & \leq & 1/2^r \nonumber
\end{eqnarray}

because $m \geq r$ implies $n/(n+m-r) \leq 1$ and, for all $0 \leq k $, $(m-k)/(n+m-k) \leq 1/2$.
\enproof

Let $P_b$ and $S_b$ be the $P$ and $S$ lists, respectively, at the bottom of the \textbf{while} loop,
after all rearrangements and adjustments to the lists.
See Figure \ref{sublist1}.
\begin{lemma}
\label{PrlenP}
$Pr(|P_b| = p) \leq  1/2^{p-1}$.
\end{lemma}
\noindent
\textbf{Proof}
\ \ \\
Suppose there are $n$ elements of type \emph{left} and $m$ of type \emph{right} distributed across $S_b$ 
and $P_b$.
Then $m$ is within one of $n$ because the numbers of each type remaining in $Sh$ are within one
of each other and the input lists were of equal length.
The number of possible merged arrangements of these $m+n$ elements is $(m+n)!/(m!n!)$.

Without loss of generality, suppose the elements in $P_b$ are of type \emph{left}.
Then the number of arrangements consistent with the existence of $|P_b| = p$ elements all greater than any
element in $S_b$ is the number of ways that $S_b$ can result from the merge of $n$ with $m-p$
elements, i.e.,  $(m+n-p)!/((m-p)! n!)$.
Hence the probability that $|P_b| = p$ is given by:
\begin{eqnarray}
Pr(|P_b| = p) &=& \frac{(m+n-p)!} {(m-p)! n!} \times \frac{m! n!}{(m+n)!} \nonumber \\
      &=& \frac{ m(m-1)(m-2)\cdots (m-p+1)} {(m+n)(m+n-1)(m+n-2) \cdots (m+n-p+1)} \nonumber \\
      &=& \frac{m}{(m+n)} \times \frac{m-1}{m+n-1} \times \frac{m-2}{m+n-2} \cdots \times \frac{m-p+1}{m+n-p+1} 
           \nonumber \\
      & \leq & 1/2^{p-1} \nonumber
\end{eqnarray}
because $m/(m+n) < 1$ and, for all $0 \leq k < n$, $(m-k)/(m+n-k) \leq 1/2$. 
\enproof

Lemmas \ref{Prscan} and \ref{PrlenP} allow us to show that the expected time complexity of all the 
loop procedures is a constant.

\begin{lemma}
\label{tscan}
The expected time used by the Scan procedure is constant.
\end{lemma}
\noindent
\textbf{Proof}
\ \ \\
The time taken to scan $2r$ elements is $k_1r$, for some constant $k_1$.
Hence the expected time, \emph{expt-scan}, is given by:
\begin{eqnarray}
\emph{expt-scan} &=& \sum_r k_1rPr(|D|=2r) \leq \sum_{r=1}^\infty k_1r/2^r = 2k_1, \nonumber
\end{eqnarray}
by Lemma \ref{Prscan}.
\enproof

Because element comparisons are restricted to the Scan procedure, Lemma \ref{tscan} tells us that the
average number of comparisons per loop traversal is a constant.
\begin{lemma}
\label{tunshuff}
The expected time used by the unShuffle procedure is constant.
\end{lemma}
\noindent
\textbf{Proof}
\ \ \\
The time required to Unshuffle a list $D$, where $|D| = 2r$, is $k_2r$, for some constant $k_2$.
See Section 4 below.
Hence the expected time, \emph{expt-shuff}, is given by:
\begin{eqnarray}
\emph{expt-shuff} &=& \sum_r k_2rPr(|D|=2r) \leq \sum_{r=1}^\infty k_2r/2^r = 2k_2, \nonumber
\end{eqnarray}
by Lemma \ref{Prscan} and the fact that the unshuffle works on the result of the scan.
\enproof
\newpage
\begin{lemma}
\label{trotate}
The expected time used by the Rotate procedure is constant.
\end{lemma}
\noindent
\textbf{Proof}
\ \ \\
$P_b+1$ elements are rotated, where $P_b$ is list $P$ at the bottom of
the loop, i.e., after rotation.
The time taken to rotate $|P_b|+1$ elements is $\leq (|P_b|+1) k_3$, for some constant $k_3$.
Hence the expected time, \emph{expt-rot}, is given by:
\begin{eqnarray}
\emph{expt-rot} &=& \sum_p (p+1)k_3Pr(|P_b| = p)\leq \sum_{p=1}^\infty (p+1)k_3/2^p = 
    \sum_{p=1}^\infty pk_3/2^p +  \sum_{p=1}^\infty k_3/2^p = 3k_3, \nonumber
\end{eqnarray}
by Lemma \ref{PrlenP}.
\enproof

Because element moves are restricted to the Unshuffle and Rotate procedures, Lemmas \ref{tunshuff} and \ref{trotate}
tell us that the average number of moves per loop traversal is a constant.

\begin{theorem}
\label{avgtime}
The average time complexity of the algorithm is $O(n)$, where $n$ is the combined length of the
two input lists.
\end{theorem}
\noindent
\textbf{Proof}
\ \ \\
Consider the merge procedures.
The actions inside the loop are either constant time operations or scans, rotations or unshuffles
which, by Lemmas \ref{tscan}, \ref{tunshuff}, \ref{trotate} and the ``linearity of expectations''
\cite[Appendix C]{cormen:alg:2001}, are expected constant time operations.
Hence the expected time to traverse the \textbf{while} loop is a constant.

Now consider the general case where the problem is broken down to the merge of a sequence of equal length
lists of lengths say $n_1$, $n_2 \ldots  n_k$.
We observe that \(\sum_{i=1}^k n_i = n\).
Since each merge takes time $O(n_i)$, the total time is $O(n)$.
\enproof
\end{document}